\documentclass[11pt,a4paper]{article}
\usepackage{amsfonts,amssymb}
\usepackage{cite}
\usepackage[T2A]{fontenc}
\usepackage[cp1251]{inputenc}
\usepackage[english]{babel}
\usepackage{amsmath}%
\usepackage[unicode]{hyperref}

\def\eq#1{\begin{equation}#1\end{equation}}
\def\matrixx#1{\left(\begin{array}{cc}#1\end{array}\right)}

\def\eqs#1{\begin{equation}\begin{split}#1\end{split}\end{equation}}

\title{\bf On a direct algorithm for constructing recursion \\ operators and Lax pairs for integrable models}
\author{\bf I.T. Habibullin and   A.R. Khakimova}

\begin{document}
\maketitle



\abstract{We suggested an algorithm for searching the recursion operators for nonlinear integrable equations. It was observed that the recursion operator $R$ can be represented as a ratio of the form  $R=L_1^{-1}L_2$ where the linear differential operators $L_1$ and $L_2$ are  chosen in such a way that the ordinary differential equation $(L_2-\lambda L_1)U=0$ is consistent with the linearization of the given nonlinear integrable equation for any value of the parameter $\lambda\in \textbf{C}$. For constructing the operator $L_1$ we use the concept of the invariant manifold which is a generalization of the symmetry. Then for searching $L_2$ we take an auxiliary linear equation connected with the linearized equation by the Darboux transformation. Connection of the invariant manifold with the Lax pairs and the Dubrovin-Weierstrass equations is discussed.}

\large

\section{Introduction}

In a series of our works we have suggested a method for construction the recursion operators and the Lax pairs for the nonlinear integrable equations (see  \cite{HabKhaPo}-\cite{{HabKhaJPA17}}). Let us give a brief explanation of core of the method. For the sake of convenience we first take a class of evolutionary type nonlinear PDE although as it is shown below the algorithm can be applied to any kind of the integrable equations.
Let us consider an integrable equation of the form
\begin{equation} \label{eq0}
u_t =f(u,u_1,u_2,...,u_k), \qquad \mbox{where} \quad u_j=D_x^{j} u.
\end{equation}
Here $D_x$ stands for the operator of the total derivative with respect to the variable $x$. In what follows we use the linearization of the equation (\ref{eq0}) around its arbitrary solution $u(x,t)$:
\begin{equation}\label{eq1}
U_t=F_*U, \quad F_*=\left(\frac{\partial f}{\partial u}+\frac{\partial f}{\partial u_1}D_x+\frac{\partial f}{\partial u_2}D_x^2+...+\frac{\partial f}{\partial u_k}D_x^k\right).
\end{equation}
We look for an ordinary differential equation 
\begin{equation} \label{eq2}
H(x,t,U,U_1,\dots,U_m;u,u_1,\dots,u_{m_1})=0
\end{equation}
of the order $m\geq 1$ compatible with the linearized equation (\ref{eq1}) for all values of the dynamical variables  $u,u_1,u_2,...$ considered here as parameters. 
The compatibility condition of the equations (\ref{eq1}) and (\ref{eq2}) coincides with the equation
\begin{equation} \label{eq3}
D_t H(x,t,U,U_1,\dots,U_m;u,u_1,\dots,u_{m_1})=0 \quad\mbox{mod} \,(1),\, (2),\, (3).
\end{equation}
It is supposed that in (\ref{eq3}) all of the derivatives are expressed due to the equations (\ref{eq0}) and (\ref{eq1}) and the variables $U_{s+m}$ with $s\geq 0$ are expressed due to the equation (\ref{eq2}).

In this case we say that  equation (\ref{eq2}) defines an invariant manifold in the space of the variables $U,\,U_1,\,U_2,\dots$. By the construction equations (\ref{eq1}) and (\ref{eq2}) are compatible if $u=u(x,t)$ is a solution of the equation (\ref{eq0}). It is remarkable that for an appropriate choice of the equation (\ref{eq2}) the converse is also true: the compatibility of the equations (\ref{eq1}) and (\ref{eq2}) implies (\ref{eq0}). As it is commonly known just this property is principal for  the Lax pairs.
Studied examples convince that the ODE (\ref{eq2}) might be of two different kinds. Namely function $H$ defining equation (\ref{eq2}) is either linear or quadratic 
\begin{eqnarray}
&& H=\sum_{j=0}^{j=m}\alpha_j(u,u_1,\dots)U_j,\label{linear}\\
&& H=c+\sum_{i,j=0}^{j=s}\alpha_{i,j}(u,u_1,\dots)U_iU_j.\label{quadratic}
\end{eqnarray}
Here $c$ is an arbitrary constant. We observed that the former case is connected with the recursion operator $R$ for the equation (\ref{eq0}) while the latter corresponds to the usual Lax pair for $c=0$ and a nonlinear Lax pair for $c\neq0$ from which the well known Dubrovin-Weierstrass equations are derived connected with the finite-gap integration procedure. 

Recall that the recursion operator is a solution of the equation (\ref{generalrecursion}) (see below). In general it is a pseudo-differential (pseudo-difference for the lattices) operator generating symmetries of the equation (\ref{eq0}). The Lax pair and the recursion operator are very close to each other. As it follows from (\ref{generalrecursion}) the operators $R$ and $\frac{d}{dt}-F_*$ generate a Lax pair:
\begin{equation} \label{eq4}
R\Psi=\lambda\Psi, \quad \frac{d}{dt}\Psi=F_*\Psi
\end{equation}
for the equation (\ref{eq0}) (see \cite{ShabatIbragimov,Sul}). Moreover, as it is demonstrated in \cite{HabKhaJPA17,HabKhaTMP17} by numerous examples the Lax pair (\ref{eq4}) is effectively converted into the classical one. 
A clear expression of this circumstance is that a linear invariant manifold is transformed into a nonlinear one by decreasing order in the corresponding ordinary differential equation (\ref {eq2}).

Note that there is a great variety of approaches for searching the Lax pairs from the Zakharov-Shabat dressing \cite{ZakharovShabat74,ZakharovShabat79} and the method of pseudopotentials  by Wahlquist and Estabrook \cite{WahlquistEstabrook}  to the singular manifold method \cite{Weiss, Conte} of Painlev$\acute{e}$ analysis and 3D consistency approach developed in \cite{NijhoffWalker,BobenkoSuris,Nijhoff}. We mention also approaches proposed in  \cite{Yam82, Xenitidis, ShabatIbragimov}. We stress that our method uses a close idea but differs from the well-known method of  \cite{WahlquistEstabrook}, where both of the Lax equations are assumed to be linear and unknown. In contrast we for a given integrable equation take its linearization as one of the Lax equations and look for the second one which is not assumed to be linear. Actually at the first stage we find a nonlinear Lax pair and then linearize it by an appropriate point transformation. Since we search only one of two equations our method is rather effective.

The problem of finding the recursion operator for an integrable equation is not easily solved. There are several methods to study the task. Some of them use the Lax representation (see, for instance, \cite{Gelfand,Magri,Sokolov,Gurses,Gardner,ShabatIbragimov}).  Others are based directly on studying the defining equation (\ref{generalrecursion}). To solve this equation the most authors use the multi-Hamiltonian approach \cite{Krasilshchik,Oevel,Zhang,Gurses2,Svinolupov}. Their basic aim is to determine two Hamiltonian operators $H_1$ and $H_2$ to the given equation. Then the sought recursion operator is given by a simple formula
\begin{equation} \label{Hamilt}
R=H_2H_1^{-1}.
\end{equation}

In the present article we use an alternative representation for the recursion operator
\begin{equation} \label{Symm}
R=L_1^{-1}L_2,
\end{equation}
where $L_1$ and $L_2$ are differential (difference) operators chosen in such a way that the equations $L_1U=0$ and  $L_2U=0$ define linear invariant manifolds . Evidently two representations (\ref{Hamilt}) and (\ref{Symm}) of the recursion operator are essentially different.

Let us briefly discuss on the content of the article. In \S 2 we study the close connection between the symmetries of the given nonlinear equation and the invariant manifolds of its linearization which provides an easy way to construct linear invariant manifolds. Then in \S 3 we give an explanation of the consecutive algorithm for defining the operators $L_1$ and $L_2$ used in the formula (\ref{Symm}). In \S\S 3.1-3.4 we illustrate the application of the algorithm with examples of the KdV equation, Volterra type autonomous and nonautonomous lattices and a system of two lattices. A direct method of finding and some applications of the nonlinear invariant manifolds in the integrability theory are demostrated in \S 4.

\section{Invariant manifolds and symmetries}

We observed that there exists a close connection between symmetries and the invariant manifolds.

{\bf Proposition 1.} {\it Any set of the symmetries 
\begin{equation} \label{Symms}
u_{\tau_j}=U^{(j)},\quad j=1,2,\dots ,k
\end{equation}
of the equation (\ref{eq0}) defines an invariant manifold of the form (\ref{eq2}) for the linearized equation (\ref{eq1}). Here $U^{(j)}$ is a solution to the equation (\ref{eq1}), depending on a finite number of the dynamical variables.}

Proof. Let us consider a linear space spanned by the functions $U^{(j)}$. Then any element $U$ of the space is represented as the linear combination
\begin{equation} \label{SymmsComb}
U=c_1U^{(1)}+c_2U^{(2)}+\dots +c_kU^{(k)}.
\end{equation}
Here we suppose that the functions $U^{(j)}$ are linearly independent. Let us consider equation (\ref{SymmsComb}) as the general solution of a differential equation of the form
\begin{equation} \label{SymmDif}
L_1U:=(D_x^k+a^{(1)}D_x^{k-1}+\dots +a^{(k)})U=0.
\end{equation}
By construction $U$ solves also the linearized equation (\ref{eq1}), therefore (\ref{SymmDif}) defines a linear invariant manifold. The coefficients of this linear differential equation are evidently found from the equation
\begin{equation} \label{SymmsMat}
L_1U=\det\left( \begin{array}{cccc}
U^{(1)}_k&U^{(1)}_{k-1}&\dots &U^{(1)}\\U^{(2)}_k&U^{(2)}_{k-1}&\dots &U^{(2)} \\\dots &\dots&\dots& \dots \\ U^{(k)}_k&U^{(k)}_{k-1}&\dots &U^{(k)}\\U_k&U_{k-1}&\dots &U \end{array} \right),
\end{equation}
where $U_m=D^m_xU$ and $U^{(j)}_m=D^m_x U^{(j)}$. The proof is complete.

On the other side for any common solution $U$ of the equations (\ref{eq1}) and (\ref{eq2}) the equation $u_{\tau}=U$ defines a symmetry (possibly nonlocal) for the equation (\ref{eq0}).

\section{Algorithm for constructing linear invariant manifolds and the recursion operators for integrable equations}

Invariant manifolds can be found directly by solving the defining equation (\ref{eq2}) which is highly overdetermined and usually effectively solved. However this way becomes rather labor consuming when applied to the  systems or higher order equations. 

To look for a linear invariant manifold we suggest a novel convenient algorithm. We observed that the linear invariant manifold is closely connected with the recursion operator $R$ of the equation (\ref{eq0}). Let us consider the equation of the form  
\begin{eqnarray}
R U = \lambda U \label{rec}
\end{eqnarray} 
and then bring this equation by multiplying from the left by a differential operator $L_1$  to the form
\begin{eqnarray}
L_2 U = \lambda L_1U \label{rec-inv}
\end{eqnarray} 
where $L_2$ is also a differential operator. Evidently the operator $L_1$ is not unique. We require that the order of $L_1$ is as small as possible. Then equation (\ref{rec-inv}) defines an invariant manifold for arbitrary value of $\lambda$ including $\lambda=0$ and $\lambda=\infty$. Therefore we can easily obtain a linear invariant manifold for the known recursion operator. However, we are going to apply this scheme in the opposite direction.

Here we use an assumption that the recursion operator admits the representation $R=L_1^{-1}L_2$ as a ratio of two differential operators defining the invariant manifolds like $L_1U=0$ and $L_2U=0$.
It is well known that usually the recursion operator satisfies the following equation \cite{Olver}
\begin{equation}\label{generalrecursion}
\frac{d}{dt}R=[F_*,R].
\end{equation}
Let us replace $R=L_1^{-1}L_2$ in (\ref{generalrecursion}) and get after a slight transformation
\begin{equation}\label{temp} 
\frac{d}{dt}L_2=\left(\frac{d}{dt}(L_1)L_1^{-1}+L_1F_*L_1^{-1}\right)L_2-L_2F_*
\end{equation}
where $F_*$ is the linearization operator (\ref{eq1}). Operator $L_1$ can easily be found due to the formula (\ref{SymmsMat}) above through the symmetries of the equation (\ref{eq0}). Let us define a new operator $A$ according to the equation 
\begin{equation}\label{generalA}
A:=\frac{d}{dt}(L_1)L_1^{-1}+L_1F_*L_1^{-1}.
\end{equation}
Then (\ref{temp}) immediately gives rise to the equation 
\begin{equation}\label{generalL2} 
\frac{d}{dt}L_2=AL_2-L_2F_*
\end{equation}
which allows to find the operator $L_2$. As a result we obtain the recursion operator  $R=L_1^{-1}L_2$ solving the equation (\ref{generalrecursion}).

{\bf Remark}. Equations (\ref{generalA}) and (\ref{generalL2}) generate two Darboux-type transformations $U\rightarrow V=L_1U$ and $U\rightarrow V=L_2U$ relating linearized equation (\ref{eq1}) with one and the same linear equation $V_t=AV$.

Below we approve the efficiency of the algorithm with several examples.

\subsection{Illustration with the KdV equation}

Consider the KdV equation
\begin{equation} \label{kdv_eq}
	u_t=u_{xxx}+uu_x.
\end{equation}
Its linearization looks as follows
\begin{equation} \label{lin_kdv}
	U_t=U_{xxx}+uU_x+u_xU.
\end{equation}
In this case $L_1$ can be defined by the formula (\ref{SymmsMat}) from Proposition 1 with $k=1$ and with the symmetry $U^{(1)}=u_x$. We set $L_1=D_x\frac{1}{u_x}$ since this operator evidently annulates $u_x$.

Let us define the operator $A$ due to the formula  
\begin{equation} \label{A}
A=D_t(L_1)L_1^{-1}+L_1F_*L_1^{-1}
\end{equation}
where $F_*=D_x^3+uD_x+u_x$ is the linearization operator and evidently $L_1^{-1}=u_xD_x^{-1}$. Equation (\ref{A}) implies
\begin{equation} \label{AA}
A=D_x^3+ \left(u-3\left(\frac{u_{xxx}}{u_x}- \frac{u_{xx}^2}{u_x^2}\right)\right)D_x+2 \frac{uu_{xx}}{u_x}-3 \frac{u_{xx}u_{xxx}}{u_x^2}-3 \frac{u_{xx}^3}{u_x^3}.
\end{equation}
Now we have to look for an operator of the order $m>1$:
\begin{equation} \label{L2}
L_2=\alpha D^{m}_x+\alpha_1D_x^{m-1}+\cdots+\alpha_m
\end{equation}
solving the following equation
\begin{equation} \label{L2L2}
D_t(L_2)=AL_2^{-1}+L_2F_*.
\end{equation}
It is easily proved that for $m=2$ equation (\ref{L2L2}) has no solution of the form (\ref{L2L2}). For $m=3$ we have
\begin{equation} \label{L2found}
L_2= D^3_x - \frac{u_{xx}}{u_x} D^2_x + \frac{2 u}{3} D_x + u_x - \frac{2 u u_{xx}}{3 u_x}.
\end{equation}
Direct computations based on the equalities (\ref{A}), (\ref{L2L2}) convince that the operator $R=L_1^{-1}L_2$ solves the equation $D_t(R)=[F_*,R]$ and therefore $R$ is the recursion operator for the KdV equation. Indeed we find from the formula $R=L_1^{-1}L_2$ that (cf. \cite{Gardner})
$$R=D_x^2+\frac{2}{3}u+\frac{1}{3}u_xD_x^{-1}.$$

\subsection{Evaluation of the invariant manifolds and the recursion operators for non-autonomous Volterra type integrable lattices}

Due to the articles \cite{Yam,Levi} we know that integrable lattices of the form 
\begin{equation}\label{volterratype}
u_{n,t}=f(n,u_{n+1},u_n,u_{n-1})
\end{equation}
satisfy a sequence of the integrability conditions. Below we mention some of these conditions which have important applications in the symmetry classification 
\begin{eqnarray}
&& D_t\log \frac{\partial f}{\partial u_1}=(D-1)q^{(1)},\label{ccl1}\\
&& D_t\log \frac{\partial f}{\partial u_{-1}}=(D-1)q^{(-1)}, \label{ccl2}\\
&& r^{(k)}=(D-1)s^{(k)},\quad k=1,2,\label{ccl3}
\end{eqnarray}
where $D$ is the shift operator, such that $Dy(n)=y(n+1)$ and $q^{(1)}$, $q^{(-1)}$, $s^{(1)}$, $s^{(2)}$ are some functions depending on  the dynamical variables $u_n$, $u_{n\pm 1}$, $u_{n\pm 2}$, \dots. In the formulas above we used the following commonly accepted abbreviated notations, which are used also everywhere below: $u:=u_{n},\,u_1:=u_{n+1},\, u_{-1}:=u_{n-1}$ and so on. Here 
\begin{eqnarray}
&&r^{(1)}=\log \left\{-\frac{f_{u_1}}{f_{u_{-1}}}\right\}, \label{r1}\\
&&r^{(2)}=s_{t}^{(1)}+2f_u \label{r2}.
\end{eqnarray}

The problem of the complete classification for autonomous lattices of the form (\ref{volterratype}) has been solved years ago (see \cite{Yam82}). Recursion operators for these equations in the autonomous case are discussed in \cite{Mikh}. Below we show that the method of invariant manifolds can be successfully applied to the integrable lattices in both autonomous and non-autonomous case. Moreover, we derive a general formula for the recursion operator which fits for a subclass of the integrable lattices (\ref{volterratype}).

First we find the linearization of the equation (\ref{volterratype})
\begin{equation}\label{linvolterratype}
U_t=\frac{\partial f}{\partial u_1}U_1+\frac{\partial f}{\partial u}U+\frac{\partial f}{\partial u_{-1}}U_{-1}.
\end{equation}

Notion of the invariant manifold introduced above for the evalutionary type PDE is easily adopted to other classes of equations.

We say that an ordinary difference equation of the form 
\begin{equation}\label{difference}
H(n,t,U_{m},\dots U_{m_1};u_{l},\dots u_{l_1})=0
\end{equation}
where $U=U(n,t)$ is considered as an unknown function and the variables $u_j$ are considered as parameters, determines an invariant manifold for the linearized equation (\ref{linvolterratype}) if the equations  (\ref{linvolterratype}), (\ref{difference}) are consistent for any choice of the solution $u=u(n,t)$ of the equation (\ref{volterratype}).

The simplest invariant manifold is found very easily, to this end we use the classical symmetry $u_\tau=cu_t$ of the equation (\ref{volterratype}). Evidently the function $U=cu_t$ determines the general solution of the linear difference equation $\displaystyle{(D-1)\frac{1}{u_t}U=0}$. Therefore equation $L_1U=0$ with $\displaystyle{L_1=(D-1)\frac{1}{u_t}}$ defines an invariant manifold for the linearized equation (\ref{linvolterratype}).

Let us first consider only those equations for which the operators $L_1$ and $L_2$ are of the form 
\begin{equation}\label{simpleL1L2}
L_1=(D-1)\frac{1}{u_t} \quad \mbox{and}\quad L_2=\alpha D^2+\beta D+\gamma+\delta D^{-1} 
\end{equation}
where the coefficients $\alpha, \beta, \gamma, \delta$ are functions of the dynamical variables ${u_j}$.
Note that this class contains all of the lattices in the Yamilov's list except the equation V4 for which $L_1$ is the second order difference operator which annulates $u_t$ and the next symmetry of the equation V4. Introduce a notation for the linearization operator of the lattice (\ref{volterratype}) by setting $f^*=f_{u_1}D+f_{u}+ f_{u_{-1}}D^{-1}$. Operator $A$ is defined due to the formula $A:=\frac{d}{dt}(L_1)L_1^{-1}+L_1f^*L_1^{-1}$ (cf. (\ref{generalA})) 
\begin{equation}\label{volterrratypeA}
A=(D-1)\left(\frac{f_{u_1}f_1}{f}-\frac{f_{u_{-1}}f_{-1}}{f}D^{-1}\right).
\end{equation}
Now we look for the operator $L_2$ from the following equation 
\begin{equation}\label{volterratypeL2} 
\frac{d}{dt}L_2=AL_2-L_2f^*.
\end{equation}

Let us rewrite (\ref{volterratypeL2}) in an explicit form 
\begin{eqnarray}\label{volterratypeL2L2}
\alpha_t D^2+\beta_t D+\gamma_t+\delta_t D^{-1}&&=(D-1)\left(\frac{f_{u_1}f_1}{f}-\frac{f_{u_{-1}}f_{-1}}{f}D^{-1}\right)\left(\alpha D^2+\beta D+\gamma+\delta D^{-1}\right)\nonumber  \\
  &&- (\alpha D^2+\beta D+\gamma+\delta D^{-1})(f_{u_1}D+f_{u}+ f_{u_{-1}}D^{-1}).\label{+6}
\end{eqnarray}

By comparing the coefficients in front of $D^3$ and $D^{-2}$ in this equation we find
\begin{equation}\label{+7}
\alpha=\left( \frac{f_{u_1}}{f}\right)_1,\quad \delta=\frac{f_{u_{-1}}}{f}.
\end{equation}
Recall that $h_1$ means $D(h)$.  Comparison of the coefficients at $D^2$ gives rise to the equation
\begin{equation}\label{volterrratypeD2}
\alpha_t=\left(\frac{f_{u_1}f_1}{f}\right)_1\beta_1-\left(\frac{f_{u_1}f}{f_1}+ \frac{(f_{u_{-1}})_1f}{f_1} \right)\alpha-\alpha(f_u)_2-\beta(f_{u_1})_1.
\end{equation}
Now we divide (\ref{volterrratypeD2}) by $\alpha$ and get
$$ D_t\log \alpha=(D-1)\beta f_1- \left(\frac{f_{u_1}f_1}{f}+ \frac{{(f_{u_{-1}})_1}f}{f_1} \right)-(f_u)_2.$$
Let us evaluate l.h.s. of the last equation due to (\ref{+7}) and find

\begin{equation}\label{intermediate}
D_t\log \alpha=D(D_t\log f_{u_1}-D_t\log f).
\end{equation}
By combining the equations  (\ref{volterrratypeD2}), (\ref{intermediate}) and replacing due to (\ref{ccl1}) we obtain the equation $(D-1)(\beta f_1 -q^{(1)}_1+\frac{f_{u_1}f_1}{f}-(f_u)_1)=0$ which implies

\begin{equation}\label{2beta}
\beta= \frac{1}{f_1}q^{(1)}_1-\frac{f_{u_1}}{f}-\frac{(f_u)_1}{f_1}.
\end{equation}

In a similar way by gathering the coefficients at $D^{-1}$ and then applying the conservation law (\ref{ccl2}) one can derive an explicit expression for $\gamma$:
\begin{equation}\label{gamma}
\gamma= -\frac{1}{f}q^{(-1)}_1+\frac{f_{u}}{f}-\left(\frac{(f_{u_{-1}})_1}{f_1}\right).
\end{equation}

Thus all the coefficients of the sought operator $L_2$ are found, however to approve that equation (\ref{volterratypeL2L2}) is completely satisfied we have to check the last two equations which are obtained by collecting the coefficients in front of $D$ and $D^0$

\begin{equation}\label{3beta}
\beta_t= \left(\frac{f_{u_1}f_1}{f}\right)_1\gamma_1-\left(\frac{f_{u_1}f_1}{f}+ \frac{(f_{u_{-1}})_1f}{f_1} \right)\beta
+\frac{f_{u_{-1}}f_{-1}}{f}\alpha_{-1}
-\alpha(f_{u_{-1}})_2-\beta(f_{u})_1 -\gamma f_{u_1},
\end{equation}

\begin{equation}\label{3gamma}
\gamma_t= \left(\frac{f_{u_1}f_1}{f}\right)_1\delta_1-\left(\frac{f_{u_1}f_1}{f}+ \frac{(f_{u_{-1}})_1f}{f_1} \right)\gamma
+\frac{f_{u_{-1}}f_{-1}}{f}\beta_{-1}
-\beta(f_{u_{-1}})_1-\gamma(f_{u}) -\delta (f_{u_1})_{-1}.
\end{equation}

Assume the coefficients $\alpha$, $\beta$, $\delta$, and $\gamma$ found above to satisfy equations (\ref{3beta}) and (\ref{3gamma}). Then the operator $R=L_1^{-1}L_2$:
\begin{equation}\label{rec2}
R=f(D-1)^{-1}(\alpha D^2+\beta D+\gamma+\delta D^{-1})
\end{equation}
is the recursion operator for the lattice (\ref{volterratype}). Indeed, it is easily verified that the found $R$ solves the following equation
\begin{equation}\label{rec-eq}
R_t=[f^*, R]
\end{equation}
which is nothing else but the defining equation for the recursion operator (see the survey \cite{Yam}).

Let us take the Volterra chain 
\begin{equation}\label{examplevolterra}
u_t=u(u_1-u_{-1})
\end{equation}
as an illustrative example, then the operators $L_1$ and $L_2$ are as follows
\begin{eqnarray}
L_1&=&(D-1)\frac{1}{u(u_1-u_{-1})},\nonumber \\
L_2&=&\frac{1}{u_2-u}D^2+\left(\frac{u_2+u_1}{u_1(u_2-u)}-\frac{1}{u_1-u_{-1}}\right)D+\frac{1}{u_2-u}- \label{volterraL12}\\
&&-\frac{u+u_{-1}}{u(u_1-u_{-1})}-\frac{1}{u_1-u_{-1}}D^{-1}.\nonumber
\end{eqnarray}
Therefore the formula $R=L^{-1}_1L_2$ gives the well known recursion operator  
\begin{equation}\label{RecVolterra}R=uD+u+u_1+uD^{-1}+u(u_1-u_{-1})(D-1)^{-1}\frac{1}{u}\end{equation}
found years ago in \cite{{Zhang}}. Note that the operators $L_1$ and $L_2$ differ from the Hamiltonian operators $H_1$, $H_2$ used in \cite{{Mikh}} to represent the recursion operator (\ref{RecVolterra}) as a ratio $R=H_2H_1^{-1}$ where
$$H_1=u(D-D^{-1})u, \quad H_2=u(DuD+uD+Du-uD^{-1}- D^{-1}u-D^{-1}uD^{-1})u.$$

We studied examples of the lattices above under assumption that $L_1$ is the first order operator (see (\ref{simpleL1L2}) above). However for some lattices this assumption is too restrictive, such that the needed $L_2$ does not exist. Then we take $L_1$ of the order higher than one. The Proposition 1 discussed above can be applied to the case of lattices as well. 

{\bf Proposition 2.} {\it Any set of the symmetries 
\begin{equation} \label{SymmsD}
u_{\tau_j}=U^{(j)},\quad j=1,2,\dots ,k
\end{equation}
of the equation (\ref{volterratype}) defines an invariant manifold of the form (\ref{difference}) for the linearized equation (\ref{linvolterratype}). Here $U^{(j)}$ is a solution to the equation (\ref{linvolterratype}), depending on a finite number of the dynamical variables.}

We omit the proof of Proposition 2 since it is almost verbatim repeat of the proof  of Proposition 1. The searched invariant manifold is defined by the following equation
\begin{equation} \label{SymmsMatD}
L_1U=\det\left( \begin{array}{cccc}
U^{(1)}_k&U^{(1)}_{k-1}&\dots &U^{(1)}\\U^{(2)}_k&U^{(2)}_{k-1}&\dots &U^{(2)} \\\dots &\dots&\dots& \dots \\ U^{(k)}_k&U^{(k)}_{k-1}&\dots &U^{(k)}\\U_k&U_{k-1}&\dots &U \end{array} \right)=0,
\end{equation}
where $U_m=D^mU$ and $U^{(j)}_m=D^mU^{(j)}$.

\subsection{Example of a recursion operator for the non-autonomous lattice}

In this section we consider a non-autonomus lattice of the form
\begin{equation}\label{nonaut_eq}
u_t=\frac{(-1)^{n+m}u(u^2-1)(u_1u_{-1}+1)}{ww_1}, \quad w=uu_{-1}-u_{-1}+u+1
\end{equation}
found in \cite{{GGH}} as a symmetry in the direction of $n$ of the quad equation 
\begin{equation}\label{quad_nonaut_eq}
u_{n+1,m+1}u_{n,m}(u_{n+1,m}-1)(u_{n,m+1}+1)+(u_{n+1,m}+1)(u_{n,m+1}-1)=0
\end{equation}
derived in \cite{{Hydon}}.
Here our goal is to find the recursion operator for the lattice (\ref{nonaut_eq}) according to the scheme above. Let us set for the simplicity $m=0$. We first find the linearization of (\ref{nonaut_eq})
\begin{equation}\label{nonaut_eq_linform}
U_t=f^*U
\end{equation}
where $f^*$ is the Frech$\acute{e}$t derivative of $f$
\begin{eqnarray}\label{nonaut_eq_linop}
&f^*=(-1)^{n+1}\frac{u(u^2-1)}{w_1}D+\\
&(-1)^n\frac{u_1u_{-1}+1}{ww_1}\left(3u^2-1-\frac{u(u^2-1)(u_1-1)}{w_1}-\frac{u(u^2-1)(u_{-1}+1)}{w_{-1}}\right)+(-1)^n\frac{u(u^2-1)}{w^2}D^{-1}.\nonumber
\end{eqnarray}
We take the operator $L_1$ as in (\ref{linvolterratype}) and then find $A$ due to the formula (\ref{simpleL1L2}):
\begin{equation}\label{nonaut_eq_A}
A=A^{(1)}D+A^{(0)}+A^{(-1)}D^{-1}
\end{equation}
where
\begin{eqnarray}\label{nonaut_eq_Acoeff}
&&A^{(1)}=(-1)^{n+1}\frac{u_2(u^2_2-1)(u_3u_2+1)w_1}{(u_2u+1)w^2_2w_3}, \nonumber\\
&&A^{(0)}=(-1)^{n+1}\frac{u(u^2-1)(u_1u_{-1}+1)w_2}{(u_2u+1)w^2_1w}+(-1)^{n+1}\frac{u_1(u^2_1-1)(u_2u+1)w}{(u_1u_{-1}+1)w^2_1w_2}, \\
&&A^{(-1)}=(-1)^{n+1}\frac{u_{-1}(u^2_{-1}-1)(uu_{-2}+1)w_1}{(u_1u_{-1}+1)w^2w_{-1}}.\nonumber
\end{eqnarray}
The next step is to look for the operator $L_2$ from the equation (\ref{volterrratypeA}). We find the coefficients of $L_2$ by using the explicit formulas (\ref{volterratypeL2L2}), (\ref{intermediate}) and the first canonical conservation laws (\ref{volterratype}), (\ref{ccl1}). As a result we obtain
\begin{eqnarray}\label{nonaut_eq_L2}
&&L_2=-\frac{w_1}{w_2(uu_2+1)}D^2-\left(\frac{2(u+1)}{(u_1+1)(uu_2+1)}+\frac{w_1w_2}{2u(u_1^2-1)(uu_2+1)}+\right.\nonumber\\
&&\left.\frac{uu_1-1}{u_1(u_1^2-1)(uu_2+1)}-\frac{2u_1}{u_1^2-1}-\frac{1}{u_1(uu_{-1}+1)}+\frac{u_2-1}{w_2}\right)D+\nonumber\\
&&\left(\frac{2(u_1-1)}{(u-1)(u_1u_{-1}+1)}-\frac{ww_1}{2u(u^2-1)(u_1u_{-1}+1)}-\frac{(u_1u+1)}{u(u^2-1)(u_1u_{-1}+1)}+\right.\\
&&\left.\frac{2u}{u^2-1}+\frac{1}{u(u_2u+1)}-\frac{u_{-1}+1}{w}\right)+\frac{w_1}{(u_1u_{-1}+1)w}D^{-1}.\nonumber
\end{eqnarray}
The important step is to verify whether the corresponding equations of the form (\ref{3beta}), (\ref{3gamma}) are satisfied. We checked that in this case these equations hold. Thus we have the recursion operator $R=L_1^{-1}L_2:$
\begin{eqnarray}\label{nonaut_eq_R}
&&R=(-1)^{n+1}\frac{u(u^2-1)}{w_1^2}D+(-1)^{n+1}\left(\frac{u(u^2-1)(u_1-1)(u_1u_{-1}+1)}{ww^2_1}+\frac{w}{2u_{-1}w_1}-\right.\nonumber\\
&&\left.\frac{2(u_{-1}-1)(u+1)}{u_{-1}ww_1}+\frac{(u-1)(u_1u_{-1}+1)}{u_{-1}ww_1}+\frac{2u(u-1)(u_{-1}-1)(u_1u_{-1}+1)}{w_1w^2}+\right.\\
&&\left.\frac{u_{-1}-1}{u_{-1}w_1}-\frac{u+1}{u_{-1}w_1}-\frac{(u-1)(u_{-1}^2)(u^2-1)(u_1u_{-1}+1)}{2u_{-1}w_1w^2}\right)+(-1)^{n+1}\frac{u(u^2-1)}{w^2}D^{-1}-\nonumber\\
&&u_t(D-1)^{-1}\left(\frac{2(u_{-1}+1)}{w}+\frac{2(u_1-1)}{w_1}+\frac{3u^2-1}{u(u^2-1)}\right).\nonumber
\end{eqnarray}

\subsection{Recursion operator for a coupled lattice}

Let us discuss the algorithm for constructing the recursion operator with the example of a system of lattices \cite{GHYan}
\eqs{&u_{n,t}=\frac{1}{v_{n+1}}+u_n^2v_n, \\ &v_{n,t}=-\frac{1}{u_{n-1}}-u_nv_n^2. \label{A11-main}}
Linearized system can be represented as 
\eq{\frac{d}{dt}\matrixx{U_n\\V_n}=F_n^*\matrixx{U_n\\V_n}, \label{A11-lin}} where the linearization operator or the Fr$\acute{e}$chet  derivative $F_n^*$ is given by \eq{F_n^*=F_n^{(1)}D_n+F_n^{(0)}+F_n^{(-1)}D_n^{-1} \label{A11-oplin}} where \eqs{F_n^{(1)}=\matrixx{0&-\frac{1}{v_{n+1}^2}\\0&0},\quad F_n^{(0)}=\matrixx{2u_nv_n&u_n^2\\-v_n^2&-2u_nv_n},\quad F_n^{(-1)}=\matrixx{0&0\\ \frac{1}{u_{n-1}^2}&0}. \label{A11-coeffoplin}} 

In the coordinate representation the linearized equation looks as follows
\begin{equation}\label{linearizedcoupled}
U_t=-\frac{1}{v_1^2}V_1+2uvU+u^2V,\quad V_t=\frac{1}{u_{-1}^2}U_{-1}-2uvV-v^2U.
\end{equation}
In order to construct the linear operator $L_1$ we use the classical symmetry $u_\tau=cu, v_\tau=-cv$ which is connected evidently with the following solution to the linearized equation $U=cu, V=-cv$. By excluding the constant parameter $c$ we arrive at the system of equations $(D-1)\frac{1}{v}V=0$, $U=\frac{u}{v}V,$ defining an invariant manifold. Thus we have
\eqs{L_1=\matrixx{0&0\\0&\frac{1}{v_1}}D+ \matrixx{1&\frac{u}{v}\\0&\frac{-1}{v}}. \label{L1forA11}}
We find the operator $A$ from the equation $A=D_t(L_1)L_1^{-1}+L_1F^*L_1^{-1}$
\eqs{A=\matrixx{0&0\\-v_1&0}D+\matrixx{uv&\frac{-1}{v_1}\\v+\frac{1}{v_1u^2}&\frac{1}{uv_1}}+\matrixx{\frac{u}{vu_{-1}}&\frac{u}{vu_{-1}}\\ \frac{-1}{vu_{-1}^2}&\frac{-1}{vu_{-1}^2}}D^{-1}. \label{AA11}} 
Then we look for the operator $L_2=aD+b+cD^{-1}$ from the equation $D_t(L_2)=AL_2+L_2F^*$. The answer is 
\eqs{L_2=\matrixx{0&-\frac{1}{v_{1}^2}\\v_1&\frac{-2}{v_1^2u}}D+\matrixx{-uv&-u^2\\v-\frac{1}{v_1^2u}&2u}+\matrixx{\frac{u}{vu_{-1}^2}&0\\ \frac{1}{vu_{-1}^2}&0}D^{-1}. \label{L2A11}} 
Therefore the recursion operator is given by $R=L_1^{-1}L_2$. Let us write it down in an explicit form \cite{HabKhaTMP17}
\eq{R=L^++2\matrixx{-u\\v}\left(D-1\right)^{-1}\left(\frac{1}{u^2v_{1}}-v, \frac{1}{v^2u_{-1}}-u\right), \label{A11-recursion}} 
where
\eqs{L^{(+)}=\matrixx{0&-\frac{1}{v_{1}^2}\\0&0}D+\matrixx{2uv&u^2-\frac{2u}{u_{-1}v^2}\\-v^2&\frac{2}{u_{-1}v}}+\matrixx{0&0\\-\frac{1}{u_{-1}^2}&0}D^{-1}. \label{A11-L+}}

\section{Quadratic invariant manifolds, Lax pairs and Dubrovin-Weierstrass equations}

Let us illustrate the application of the nonlinear invariant manifolds for constructing the Lax pairs to integrable models with the example of the KdV equation
\begin{equation}\label{kdv2}
u_t=u_{xxx}+uu_x.
\end{equation}
By the substitution $U=W_x$ we change the linearized equation
\begin{equation}\label{kdvlinW}
U_t=U_{xxx}+uU_x+u_xU \quad \Rightarrow \quad W_t=W_{xxx}+uW_x.
\end{equation}
and find ODE compatible with the equations above (see also \cite{HabKhaJPA17}) 
\begin{equation}\label{kdvFW}
W_{xx}=F(W_x,W,u)
\end{equation}
i.e. we request that
\begin{equation}\label{kdvFWtlinWxx}
\left.\frac{d}{dt}(W_{xx})-\frac{d}{dx}(W_t)\right|_{(\ref{kdv2}),(\ref{kdvlinW}),(\ref{kdvFW})}=0.
\end{equation}
Eq. (\ref{kdvFWtlinWxx}) must satisfy identically for all values of $W$, $W_x$, $u$, $u_x$, $u_{xx}$, ...
The consistency condition is reduced to a huge equation, which splits down into 7 equations and is effectively  solved 
\begin{eqnarray*} \label{eq_found_F_compl_kdv}
 &&3F_{uu}u_xu_{xx}+(3W_xF_{Wu}+W_x+3F_{W_xu}F)u_{xx}+(3F_{Wuu}W_x+3F_{W_xuu}F+3F_{W_xu}F_u)u_x^2 \nonumber\\
 &&+F_{uuu}u_x^3+(3F_{W_x}F_{W_xu}F+2F+3F_{WWu}W_x^2-F_{W_x}W_x+3W_xF_{W_xu}F_W+6W_xF_{W_xWu}F\nonumber\\
 &&+3W_xF_{W_xW}F_u+3F_{Wu}F+3FF_{W_xW_x}F_u+3F_{W_xW_xu}F^2)u_x+F_{WWW}W_x^3+F_{W_xW_xW_x}F^3 \\
 &&+3F_{W_xW}F^2+3W_xF_{W_x}F_{W_xW}F+3W_xFF_{W_xW_x}F_W+3F_{W_x}F^2F_{W_xW_x}+3W_x^2F_{W_xWW}F\nonumber\\
 &&+3W_x^2F_{W_xW}F_W+3W_xF_{WW}F+3W_xF_{WxWxW}F^2=0.\nonumber
\end{eqnarray*}


Compare the coefficients at the independent variables $u_{xx},u_x$:

\noindent
1. $u_{xx}u_x:$
\begin{equation*}\label{1}
\frac{d^2}{du^2}F(W_x,W,u)=0 \Rightarrow F(W_x,W,u)=F_1(W_x,W)u+F_2(W_x,W)
\end{equation*}
2. $u_{xx}u:$
\begin{equation*}\label{2}
\frac{d}{dW_x}F_1(W_x,W)=0 \Rightarrow F_1(W_x,W)=F_1(W)
\end{equation*}
3. $u_{xx}:$
\begin{equation*}\label{3}
3\frac{d}{dW}F_1(W)+1=0 \Rightarrow F_1(W)=-\frac{1}{3}W+c_1
\end{equation*}
4. $u_{x}u:$
\begin{eqnarray*}\label{4}
\begin{aligned}
\frac{d^2}{dW^2_x}F_2(W_x,W)&-\frac{1}{W-3c_1}=0 \Rightarrow \\
&F_2(W_x,W)=\frac{W^2_x}{2(W-3c_1)}+F_4(W)W_x+F_3(W)
\end{aligned}
\end{eqnarray*}
5. $u_{x}:$
\begin{equation*}\label{5}
(W-3c_1)\frac{d}{dW}F_4(W)+F_4(W)=0 \Rightarrow F_4(W)=\frac{c_2}{W-3c_1}
\end{equation*}
6. $uW_x: \quad  c_2=0$ 

\noindent
7. $u:$
\begin{eqnarray*}\label{7}
\begin{aligned}
(W-3c_1)^2\frac{d^2}{dW^2}F_3(W)&+(W-3c_1)\frac{d}{dW}F_3(W)-F_3(W)=0 \Rightarrow \\ 
&F_3(W)=\frac{c_3}{W-3c_1}+c_4(W-3c_1).
\end{aligned}
\end{eqnarray*}
Finally we obtain ODE $W_{xx}=F(W_x,W,u)$, where:
\begin{equation*}\label{8}
F(W_x,W,u)=\left(-\frac{1}{3}W+c_1\right)u+\frac{W^2_x}{2(W-3c_1)}+\frac{c_3}{W-3c_1}+c_4(W-3c_1).
\end{equation*}


Parameter $c_1$ is easily removed by the shift $\bar W=W-3c_1$, therefore we set $c_1=0$. Denote $c_4=\lambda$. 
\begin{equation}
W_{xx}W-\left(\lambda-\frac{1}{3}u\right)W^2-\frac{1}{2}W^2_x-c_3=0.
\end{equation}
We have two choices $ c_3=0$ and $ c_3\neq0$. 

1) If $ c_3=0$ then we get a nonlinear Lax pair:
\begin{eqnarray}\label{9}
\begin{aligned}
&W_{xx}=-\frac{1}{3}Wu+\frac{W^2_x}{2W}+\lambda W,\\
&W_t=\left(2\lambda -\frac{2}{3}u\right)W_x-\frac{1}{3}u_xW.
\end{aligned}
\end{eqnarray}
It is linearized by taking $W=\varphi^2$ and reduces to the usual pair:
\begin{eqnarray}\label{10}
\begin{aligned}
&\varphi_{xx}=\left(-\frac{1}{6}u+\frac{1}{2}\lambda\right)\varphi,\\
&\varphi_t=\left(2\lambda -\frac{2}{3}u\right)\varphi_x-\frac{1}{6}u_x\varphi.
\end{aligned}
\end{eqnarray}

2) If $c_3\neq 0$, then we get the well-known equation, connected with the finite gap solutions (see \cite{Dubrovin}):
\begin{equation}
W_{xx}W-\left(\lambda-\frac{1}{3}u\right)W^2-\frac{1}{2}W^2_x-c_3(\lambda)=0.
\end{equation}
We assume that $W$ and $c_3=c_3(\lambda)$ are polynomials of the form:
\begin{equation}
W=\prod^{n}_{j=1}(\lambda-r_j), \qquad c_3(\lambda)=-\prod^{2n+1}_{j=1}(\lambda-e_j).
\end{equation}
Recall that functions $r_j$ satisfy the Dubrovin-Weierstrass equations
\begin{equation*}
r_{j,x}\prod_{i\neq j}(r_j-r_i)=\sqrt{2\prod^{2n+1}_{s=1}(r_j-e_s)}, j=1,...,n.
\end{equation*}
The potential $u$ is found as follows:
\begin{equation*}
u=3\sum^{2n+1}_{i=1}e_i-6\sum^{n}_{i=1}r_i.
\end{equation*}

Let us consider the Volterra lattice
\begin{equation}\label{V}
u_{n,t}=u_n(u_{n+1}-u_{n-1}).
\end{equation}
Linearization of (\ref{V})
\begin{equation}\label{linvolterra}
U_{n,t}=U_n(u_{n+1}-u_{n-1})+u_n(U_{n+1}-U_{n-1})
\end{equation}
is rather complicated, it contains $u_n$, $u_{n+1}$, $u_{n-1}$. Let us simplify it by the substitution $U_n=u_n(P_{n+1}-P_{n-1})$ which is the linearization of the substitution $\log u_n=p_{n+1}-p_{n-1}$ connecting the Volterra lattice with the equation $p_{nt}=e^{p_{n+1}-p_{n-1}}$.
Thus we get from (\ref{linvolterra}) the equation
\begin{equation}\label{mlinvolterra}
P_{n,t}=u_n(P_{n+1}-P_{n-1}),
\end{equation}
which contains only one parameter $u_n$.
Look for the invariant manifold in the form
\begin{equation}
P_{n+1}=F(P_n,P_{n-1},u_n)
\end{equation}
i.e. an ordinary difference equation with the parameter $u_n$.

The answer is as follows 
\begin{equation}\label{inva}
u_n(P_{n+1}+P_n)(P_n+P_{n-1})=\lambda P_n^2+c, \qquad c=const.
\end{equation}
Assume that functions 
$P_n$ and  $c$ are some polynomials on $\lambda$:
\begin{equation}
P_n=\prod^{m}_{j=1}(\lambda-\gamma_j(n)), \qquad c(\lambda)=-\prod^{2m+1}_{j=1}(\lambda-e_j).
\end{equation}
Then for $\gamma_j(n)$ and $u_n$ we obtain from (\ref{inva}) difference equations
\begin{equation*}
u_n\prod^{m}_{i=1}(\gamma_j(n)-\gamma_i(n+1))(\gamma_j(n)-\gamma_i(n-1))=-2\prod^{2m+1}_{i=1}(\gamma_j(n)-e_i), 
\end{equation*}
\begin{equation*}
u_n=\frac{1}{4}\sum^{2m+1}_{i=1}e_i-\frac{1}{2}\sum^{m}_{i=1}\gamma_i(n),\qquad j=1,...,m.
\end{equation*}
which are the discrete versions of the Dubrovin-Weierstrass equations (see \cite{Dubrovin}).

Assume that $c=0$ then the invariant surface takes the form 
\begin{equation}\label{V0}
u_n(P_{n+1}+P_n)(P_n+P_{n-1})=\lambda P_n^2.
\end{equation}
Now from the linearized equation (\ref{V0}) we obtain the equation 
\begin{equation}\label{V1}
P_{n,t}=u_n\left(\frac{\lambda P_n^2}{P_n+P_{n-1}}-P_n-P_{n-1}\right)
\end{equation}
describing the time evolution of $P$. Let us derive the Lax pair for the Volterra lattice. Since the equation (\ref{V0}) is homogeneous it is reasonable to set $Z=\frac{P}{P_{-1}}$ and reduce (\ref{V0}) to the discrete Riccati equation:
\begin{equation}\label{V2}
Z_{n+1}=\frac{Z_n(\frac{\lambda}{u_n}-1)-1}{Z_n+1}.
\end{equation}
We linearize (\ref{V2}) in the standard way by introducing $Z_n+1=\frac{\varphi_n}{\varphi_{n-1}}$:
\begin{equation}\label{V3}
\varphi_{n+1}=\frac{\lambda}{u_n}(\varphi_n-\varphi_{n-1}).
\end{equation}
By differentiating the equality $\frac{P_n}{P_{n-1}}=\frac{\varphi_n}{\varphi_{n-1}}-1$ we get after some transformations the following equation 
\begin{equation}\label{V4}
\frac{\varphi_{n,t}}{\varphi_n}-\frac{\varphi_{n-1,t}}{\varphi_{n-1}}=\lambda-u_n+u_{n-1}-\lambda\frac{\varphi_{n-1}}{\varphi_n}-u_{n-1}\frac{\varphi_n}{\varphi_{n-1}}.
\end{equation}
Now we assume that the time evolution of the eigenfunctions is given by a linear system of the form  $\varphi_{n,t}=a_n\varphi_n+b_n\varphi_{n-1}, \varphi_{n-1,t}=c_n\varphi_n+d_n\varphi_{n-1}$ and substitute this expression into (\ref{V4}). By comparing the coefficients before the powers of the independent variables $\varphi_{n},\, \varphi_{n-1}$ we find equations for the searched functions $a_n, b_n, c_n, d_n$: 
\begin{equation}\label{V5}
b_n=-\lambda,\, c_n=u_{n-1},\, a_n-d_n=\lambda-u_n+u_{n-1}.
\end{equation}
In addition we have $D(\varphi_{n-1,t})=\varphi_{n,t}$ which implies that $a_n=\lambda+d_{n+1}$. Summarizing the reasonings above we get:
\begin{eqnarray}\label{V6}
\varphi_{n,t}=(\lambda-u_n)\varphi_n-\lambda\varphi_{n-1},\\
\varphi_{n-1,t}=u_{n-1}\varphi_n-u_{n-1}\varphi_{n-1}.
\end{eqnarray}
Change of the variables $\varphi_n=\lambda^{n/2}\prod^{\infty}_{j=n}u_j\psi_{n-1}$, reduces the equations (\ref{V3}), (\ref{V6}) to the usual Lax pair for the Volterra equation 
\begin{eqnarray*}\label{V7}
&&\psi_{n+1}=\xi\psi_n-u_n\psi_{n-1}, \quad \xi=\sqrt{\lambda},\\
&&\psi_{n,t}=(\xi^2+u_n)\psi_n-\xi u_n\psi_{n-1}.
\end{eqnarray*}

\end{document}